\documentclass[final
    ,final            
  ]
  {aipproc}

\layoutstyle{6x9}

\begin{document}

\newcommand{\stackeven}[2]{{{}_{\displaystyle{#1}}\atop\displaystyle{#2}}}
\newcommand{\lsim}{\stackeven{<}{\sim}}
\newcommand{\gsim}{\stackeven{>}{\sim}}
\def\fig#1{{Fig.~\ref{#1}}}
\def\eq#1{{Eq.~(\ref{#1})}}


\title{DIS in AdS\footnote{Talk given by Yu.K. at the International
    Workshop on Diffraction in High-Energy Physics (Diffraction 2008)
    in La Londe-les-Maures, France, based on \cite{Albacete:2008ze}.}}

\classification{11.25.Tq, 12.38.-t, 13.60.Hb}
\keywords      {Parton Saturation, AdS/CFT Correspondence, Deep Inelastic Scattering}

\author{Javier L.\ Albacete}{
  address={ECT*, Strada delle Tabarelle 286, I-38050, Villazzano (TN), Italy}
}

\author{Yuri V.\ Kovchegov}{
  address={Department of Physics, The Ohio State University,
Columbus, OH 43210, USA}
}

\author{Anastasios Taliotis}{
  address={Department of Physics, The Ohio State University,
Columbus, OH 43210, USA}
}

\begin{abstract}
  We calculate the total cross section for the scattering of a
  quark--anti-quark dipole on a large nucleus at high energy for a
  strongly coupled ${\cal N}=4$ super Yang-Mills theory using AdS/CFT
  correspondence. We model the nucleus by a metric of a shock wave in
  AdS$_5$.  We then calculate the expectation value of the Wilson loop
  (the dipole) by finding the extrema of the Nambu-Goto action for an
  open string attached to the quark and antiquark lines of the loop in
  the background of an AdS$_5$ shock wave. We find two physically
  meaningful extremal string configurations. For both solutions we
  obtain the forward scattering amplitude $N$ for the quark
  dipole--nucleus scattering. We study the onset of unitarity with
  increasing center-of-mass energy and transverse size of the dipole:
  we observe that for both solutions the saturation scale $Q_s$ is
  independent of energy/Bjorken-$x$ and depends on the atomic number
  of the nucleus as $Q_s \sim A^{1/3}$. Finally we observe that while
  one of the solutions we found corresponds to the pomeron intercept
  of $\alpha_P = 2$ found earlier in the literature, when extended to
  higher energy or larger dipole sizes it violates the black disk
  limit. The other solution we found respects the black disk limit and
  yields the pomeron intercept of $\alpha_P = 1.5$. We thus
  conjecture that the right pomeron intercept in gauge theories at
  strong coupling may be $\alpha_P = 1.5$.
\end{abstract}

\maketitle


\section{Setting up the Problem}

\begin{figure}
\includegraphics[width=6cm]{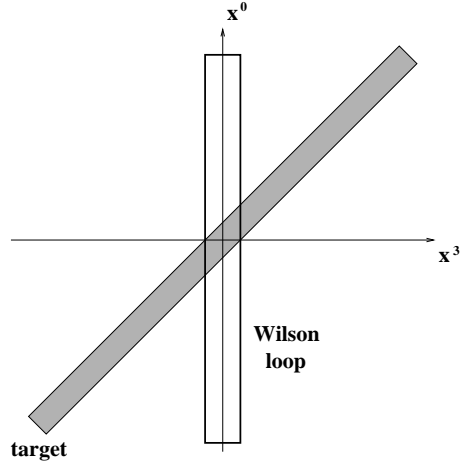}
  \caption{Dipole-nucleus scattering in the dipole's rest frame. 
    While quark and anti-quark have the same $x^3$-coordinate, we show
    them apart from each other for illustration purposes.}
  \label{rest}
\end{figure}

It is well-known that using the light-cone perturbation theory one can
decompose the total scattering cross section of deep inelastic
scattering (DIS) into a convolution of a light-cone wave function
squared ($\Phi$) for a virtual photon to decay into a $q\bar q$ pair
and the imaginary part of the forward scattering amplitude for the
$q\bar q$ pair--target interaction ($N$):
\begin{equation}\label{sigN}
  \sigma_{tot}^{\gamma* A} (Q^2, x_{Bj}) \, = \, \int \frac{d^2 r \, d
    \alpha }{2 \pi} \, \Phi ({\bf r}, \alpha, Q^2) \ d^2 b \ N({\bf
    r}, {\bf b} , Y).
\end{equation}
While $\Phi$ is well known and contains only QED interactions, the QCD
part is contained in $N({\bf r}, {\bf b} , Y)$, which is the imaginary
part of the forward scattering amplitude for the scattering of a
quark-antiquark dipole of transverse size $\bf r$ at center-of-mass
impact parameter $\bf b$ on a target, where the total rapidity of the
scattering process is $Y = \ln 1/x_{Bj}$ with $x_{Bj}$ the Bjorken $x$
variable. One can therefore relate $N({\bf r}, {\bf b} , Y)$ to the
expectation value of a fundamental Wilson loop by writing
\begin{equation}\label{N}
  N({\bf r}, {\bf b} , Y) = 1 - S({\bf r}, {\bf b} , Y)
\end{equation}
with the (real part of the) $S$-matrix
\begin{equation}\label{S}
  S({\bf r}, {\bf b} , Y) = \frac{1}{N_c} \, \mbox{Re} \, \bigg\langle
  W \left( {\bf b} + \frac{1}{2} {\bf r}, {\bf b} - \frac{1}{2} {\bf
      r}, Y \right) \bigg\rangle.
\end{equation}
Here $W \left( {\bf b} + \frac{1}{2} {\bf r}, {\bf b} - \frac{1}{2}
  {\bf r}, Y \right)$ denotes a Wilson loop formed out of a quark line
at ${\bf b} + \frac{1}{2} {\bf r}$ and an antiquark line at ${\bf b} -
\frac{1}{2} {\bf r}$ with the links connecting the two lines at plus
and minus temporal infinities, as shown in \fig{rest}. The averaging
in \eq{S}, denoted by $\langle \ldots \rangle$, is performed over all
possible wave functions of the target.

Our goal here is to find the expectation value of the Wilson loop
$\langle W \rangle$ in \fig{rest} using the AdS/CFT correspondence.
According to AdS/CFT prescription \cite{Maldacena:1998im} the
expectation value of the Wilson loop is given by the classical
Nambu-Goto action of the open string in AdS$_5$ space $\langle W
\rangle \, \sim \, e^{i S_{NG}}$ with the worldsheet of the string
connecting to the Wilson loop at the boundary of AdS$_5$. We will
model the nucleus by a smeared shock wave in AdS$_5$ with the
following metric \cite{Janik:2005zt} (note that in this approximation
there is no $\bf b$-dependence, which we will henceforth suppress in
the arguments of $S$ and $N$)
\begin{equation}\label{metricsm}
  ds^2=\frac{L^2}{z^2}\left[-2\,dx^+dx^-+\frac{\mu}{a}
      \,\theta(x^-)\,\theta(a-x^-)\,z^4\,dx^{-2}+dx_{\perp}^2+dz^2\right].
\end{equation}
Here $\mu \, = \, p^{+} \, \Lambda^2 \,A^{1/3}$ and $a \, \approx \, 2
\, R \, \frac{\Lambda}{p^+} \sim\frac{A^{1/3}}{p^+}$, where the
nucleus of radius $R$ has $A$ nucleons in it with $N_c^2$ valence
gluons each. $p^{+}$ is the light cone momentum of each nucleon and
$\Lambda$ is the typical transverse momentum scale. We thus need to
extremize the string worldsheet in the background of \eq{metricsm}.

\section{Static Solution}

One can argue that for a large enough nucleus the string configuration
should be static \cite{Albacete:2008ze}. Parameterizing the static
string as $X^{\mu}(t,x) = \left(t, x ,0 , 0, z(x) \right)$ with $x
\in\left[-\frac{r}{2},\frac{r}{2}\right]$ and $z (x = \pm r/2) \, = \,
0$ we write the Nambu-Goto action as ($\lambda$ is 't Hooft coupling)
\begin{equation}\label{sstat}
  S_{NG} (\mu) \,=\, - \, \frac{\sqrt{\lambda}}{2\,\pi}\, 
  \int\limits_{0}^{a \, \sqrt{2}} dt
  \int_{-r/2}^{r/2} dx \, \frac{1}{z^2}\,
  \sqrt{(1+z'^{\,2})\left(1-\frac{\mu}{2\,a}\,z^4\right)}.
\end{equation}
The classical string configuration extremizing the action
(\ref{sstat}) was found in \cite{Albacete:2008ze} having the shape of
a hanging string. In fact one finds six different extremal
configurations characterized by different results for the maximum of
the 5th string coordinate $z_{max}$:
\begin{equation}\label{zmax}
  z_{max} = c_0 \, r \, \sqrt{\frac{1}{3\,m\,\Delta}+\Delta},
\end{equation}
with
\begin{equation}\label{delta}
  \Delta\,=\,
  \left[-\frac{1}{2\,m}+\sqrt{\frac{1}{4\,m^2}-\frac{1}{27\,m^3}}\right]^{1/3}
  \!\exp\left[i\,\frac{2\,\pi\,n}{3}\right].
\end{equation}
We defined $\frac{\mu}{2\,a} \, = \, s^2 = p^{+ \, 2} \Lambda^2 /2$
and $m\,=\,c_0^4\,r^4\,s^2$.  The index $n$ in \eq{delta} takes on
values $n=0,1,2$, and taking principal and secondary square roots in
\eq{zmax} makes the total number of solutions for $z_{max}$ equal to
six. We pick the right solution by imposing the physical requirement
that $N (r=0, Y)=0$ (color transparency) and $N (r \rightarrow \infty,
Y) \rightarrow 1$ (black disk limit).

\vspace*{-5mm}
\section{Results and Conclusions}

\begin{figure}
\includegraphics[width=8cm]{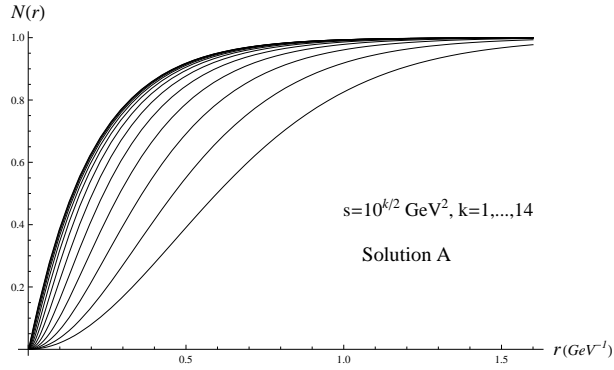}
  \caption{Dipole scattering amplitude $N(r, s)$ given by \eq{N1} 
    with $\lambda=20$, $A^{1/3}=5$ and $\Lambda = 1$~GeV.}
  \label{ndip1}
\end{figure}

The solution which naturally satisfies the above requirements is given
by Eqs. (\ref{zmax}) and (\ref{delta}) with $n=1$ and taking the
secondary square root in \eq{zmax}. Writing the result as
$z_{max}^{n=1} = - i \rho_{max} (r,s)$, using the obtained solution in
\begin{equation}\label{Smatrix}
S (r, Y) \,
= \, \mbox{Re} \left[ \frac{\langle W \rangle_\mu}{\langle W
    \rangle_{\mu \rightarrow 0}} \right] \, = \, \mbox{Re} \left[ e^{i
    \, [S_{NG} (\mu) - S_{NG} (\mu \rightarrow 0)]} \right] 
\end{equation}
to find the $S$-matrix \cite{Maldacena:1998im,Janik:1999zk}, we obtain
the dipole-nucleus scattering amplitude \cite{Albacete:2008ze} ($s
\sim e^Y$)
\begin{equation}\label{N1}
  N (r, s) \, = \, 1 - \exp \bigg\{ -
  \frac{\sqrt{\lambda}\,a}{\pi\,c_0 \, \sqrt{2}} \,
  \left[\,\frac{c_0^2\,r^2}{\rho_{max}^3 (r,s)} + \frac{2}{\rho_{max} (r,s)} -
    2 \sqrt{s} \,\right] \bigg\}.
\end{equation}
Here $c_0 \equiv \, \Gamma^2 \left(\frac{1}{4}\right) / (2 \,
\pi)^{3/2}$.  The amplitude in \eq{N1} is plotted in \fig{ndip1} as a
function of dipole size $r$ for a range of different energies $s$.

Using \eq{N1} one can easily show that at lower energies and/or small
dipole sizes it reduces to 
\begin{equation}\label{Nsmall}
  N (r, s) \bigg|_{r^2 \, s \ll 1} \, = \, 1 - \exp \bigg\{ -
  \frac{\sqrt{\lambda} \, c_0}{2 \, \pi \, \sqrt{2}} \, r^2 \, s^{1/2}
  \, \Lambda \, A^{1/3} \bigg\}.
\end{equation}
Hence for very small dipoles $N \sim s^{1/2}$. Identifying this
behavior with a single graviton exchange in the bulk, and hence with
the single pomeron exchange in the gauge theory, we obtain the pomeron
intercept of $\alpha_P - 1 = 1/2$ or, equivalently,
\begin{equation}
\alpha_P = 1.5. 
\end{equation}

\vspace*{-9mm}
At high energies \eq{N1} gives
\begin{equation}\label{Nlarge}
  N (r, s) \bigg|_{r^2 \, s \gg 1} \, = \, 1 - \exp \bigg\{ -
  \frac{\sqrt{\lambda}}{\pi \, \sqrt{2}} \, r \, \Lambda \, A^{1/3}
  \bigg\}.
\end{equation}
One can see that the amplitude $N (r, s)$ becomes energy-independent!
Defining the saturation scale $Q_s$ by requiring that $N (r = 1/Q_s,
s) = o(1)$ we get from \eq{Nlarge} $Q_s \, \sim \, \sqrt{\lambda} \,
\Lambda \, A^{1/3}$. Thus the saturation scale is independent of
energy! This conclusion appears to agree with the results of
\cite{Dominguez:2008vd}.

\begin{figure}[h]
\includegraphics[width=8cm]{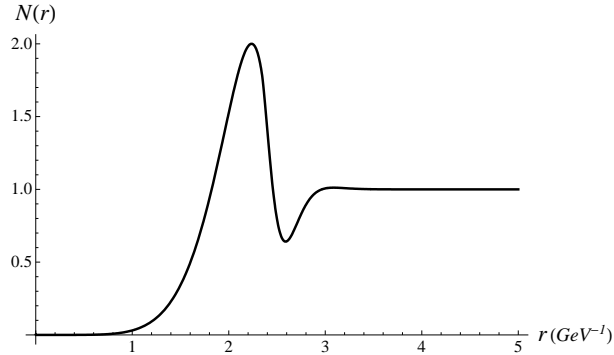}
  \caption{Dipole scattering amplitude $N(r, s)$ corresponding to 
    the string configuration given by $z_{max}^{n=2}$.}
  \label{ndip3}
\end{figure}

The only problem with the $n=1$ solution for $z_{max}$ is that it does
not map onto Maldacena's vacuum string configuration from
\cite{Maldacena:1998im} in the $\mu \rightarrow 0$ limit. While it is
not clear whether such mapping is a necessary requirement, we point
out that the $n=2$ solution for $z_{max}$ does map onto the vacuum
solution from \cite{Maldacena:1998im} in the $\mu \rightarrow 0$
limit. The dipole amplitude given by the string configuration given by
$z_{max}^{n=2}$ is plotted in \fig{ndip3}.

One can show that at lower energies and/or small dipole sizes the
dipole amplitude given by $z_{max}^{n=2}$ is (note the cosine!)
\begin{equation}\label{Nsmall2}
  N (r, s) \bigg|_{r^2 \, s \ll 1} \, = \, 1 - \cos \bigg\{ -
  \frac{\sqrt{\lambda} \, c_0^2}{2 \, \pi} \, r^3 \, s
  \, \Lambda \, A^{1/3} \bigg\}.
\end{equation}
At very small-$r$ this gives $N (r, s) \sim s^2$. Identifying this
with a double-pomeron (graviton) exchange as the small-$r$ expansion
starts from a {\em quadratic} term in the Nambu-Goto action
(diffractive dominance) we see that in this case the pomeron intercept
is given by $\alpha_P = 2$ in agreement with the results of
\cite{Brower:2006ea}. The problem with this solution is the
oscillations shown in \fig{ndip3}, which can also be seen from
\eq{Nsmall2}. Strictly speaking there is nothing wrong with $N$ going
above one (i.e. above the black disk limit) as the real constraint is
$0 \le \sigma_{inel} \propto 2 N - N^2$ leading to $N \le 2$. We note
that $N>1$ in the regions where the elastic cross section contribution
dominates in the total cross section, in agreement with
\cite{Brower:2006ea}.  However the oscillations in \fig{ndip3} appear
to have no clear physical origin and have no analogue in perturbative
calculations: this gives us a reason to doubt the validity of this
solution. In \cite{Albacete:2008ze} it is shown how a modification of
the prescription in \eq{Smatrix} leads to a more physical $N(r,s)$ for
the solution given by $z_{max}^{n=2}$ (and which is qualitatively
similar to the $n=1$ solution): however the resulting $N$ is
identically $0$ for a range of small $r$, making the determination of
the pomeron intercept impossible. To uniquely determine which one of
the $n=1$ and $n=2$ solutions is the true prediction of string theory
one probably has to calculate quantum corrections to the above
results.

\begin{figure}
  \includegraphics[width=8cm]{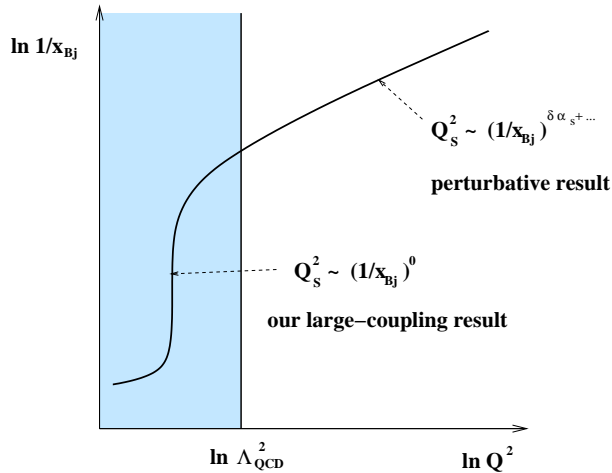}
  \caption{A sketch of the saturation line in ($\ln Q^2, \ln 1/x_{Bj}$)-plane 
    combining both perturbative (previously known) and
    non-perturbative results found in this work.}
  \label{satmap}
\end{figure}

We conclude by displaying the saturation line found in this work in
\fig{satmap}, along with a possible matching of our results on the
well-known perturbative behavior.


\vspace*{-5mm}

\begin{theacknowledgments}
\vspace*{-5mm}
  This research is sponsored in part by the U.S. Department of Energy
  under Grant No. DE-FG02-05ER41377.
\end{theacknowledgments}

\vspace*{-5mm}

\bibliographystyle{aipproc}   


\begin{thebibliography}{6}
\expandafter\ifx\csname natexlab\endcsname\relax\def\natexlab#1{#1}\fi
\providecommand{\enquote}[1]{``#1''}
\expandafter\ifx\csname url\endcsname\relax
  \def\url#1{\texttt{#1}}\fi
\expandafter\ifx\csname urlprefix\endcsname\relax\def\urlprefix{URL }\fi
\providecommand{\eprint}[2][]{\url{#2}}

\bibitem[Albacete et~al.(2008)]{Albacete:2008ze}
J.~L. Albacete, Y.~V. Kovchegov, and A.~Taliotis, \emph{JHEP} \textbf{07}, 074
  (2008), \eprint{0806.1484}.

\bibitem[Maldacena(1998)]{Maldacena:1998im}
J.~M. Maldacena, \emph{Phys. Rev. Lett.} \textbf{80}, 4859--4862 (1998),
  \eprint{hep-th/9803002}.

\bibitem[Janik and Peschanski(2006)]{Janik:2005zt}
R.~A. Janik, and R.~Peschanski, \emph{Phys. Rev.} \textbf{D73}, 045013 (2006),
  \eprint{hep-th/0512162}.

\bibitem[Janik and Peschanski(2000)]{Janik:1999zk}
R.~A. Janik, and R.~B. Peschanski, \emph{Nucl. Phys.} \textbf{B565}, 193--209
  (2000), \eprint{hep-th/9907177}.

\bibitem[Dominguez et~al.(2008)]{Dominguez:2008vd}
F.~Dominguez, C.~Marquet, A.~H. Mueller, B.~Wu, and B.-W. Xiao, \emph{Nucl.
  Phys.} \textbf{A811}, 197--222 (2008), \eprint{0803.3234}.

\bibitem[Brower et~al.(2007)]{Brower:2006ea}
R.~C. Brower, J.~Polchinski, M.~J. Strassler, and C.-I. Tan, \emph{JHEP}
  \textbf{12}, 005 (2007), \eprint{hep-th/0603115}.


\end{thebibliography}

\end{document}